\documentclass[useAMS,usenatbib]{mn2e}
\usepackage{aas_macros}
\usepackage{graphicx}
\title[Millimeter-Wave Excess Emission in RQ AGN]
{Discovery of Millimeter-Wave Excess Emission in Radio-Quiet Active Galactic Nuclei}
\author[E. Behar et al.]
{Ehud Behar$^{1}$\thanks{E-mail: behar@physics.technion.ac.il}, Ranieri D. Baldi$^{1}$,  Ari Laor$^{1}$, 
Assaf Horesh$^{2}$, 
\newauthor Jamie Stevens$^{3}$, and Tasso Tzioumis$^{3}$
\\
$^{1}$Department of Physics, Technion 32000, Haifa 32000, Israel\\
$^{2}$Weizmann Institute of Science, Rehovot, Israel\\
$^{3}$CSIRO Astronomy and Space Science, Australia}
\begin{document}


\pagerange{\pageref{firstpage}--\pageref{lastpage}} \pubyear{2014}

\maketitle

\label{firstpage}

\begin{abstract}
The physical origin of radio emission in Radio Quiet Active Galactic Nuclei (RQ AGN) remains unclear,
whether it is a downscaled version of the relativistic jets typical of Radio Loud (RL) AGN, or whether it originates from the accretion disk. 
The correlation between 5 GHz and X-ray luminosities of RQ AGN, which follows $L_R = 10^{-5}L_X$ observed also in stellar coronae, suggests an association of both X-ray and radio sources with the accretion disk corona. 
Observing RQ AGN at higher (mm-wave) frequencies, where synchrotron self absorption is diminished,
and smaller regions can be probed, is key to exploring this association.
Eight RQ AGN, selected based on their high X-ray brightness and variability,
were observed at 95~GHz with the CARMA and ATCA telescopes.
All targets were detected at the $1-10$~mJy level.
Emission excess at 95~GHz  of up to $\times 7$ is found with respect to archival low-frequency steep spectra,
suggesting a compact, optically-thick core superimposed on the more extended structures that dominate at low frequencies. 
Though unresolved, the 95 GHz fluxes imply optically thick source sizes of $10^{-4}-10^{-3}$ pc, or $\sim 10 - 1000$ gravitational radii.
The present sources lie tightly along an $L_R$ (95 GHz) = $10^{-4}L_X$ (2--10 keV) correlation, analogous to that of stellar coronae and RQ AGN at 5~GHz, while RL AGN are shown to have higher $L_R / L_X$ ratios.
The present observations argue that simultaneous mm-wave and X-ray monitoring of RQ AGN features a promising method for understanding accretion disk coronal emission. 
\end{abstract}


\section{Introduction}
Radio loud (RL) active galactic nuclei (AGN) are known for their relativistic jets that often extend beyond the nucleus.  
Radio emission from Radio Quiet (RQ) AGN \citep[as defined by][]{kellerman89} is lower by several orders of magnitude than that of RL AGN.  
RQ AGN show, on average, smaller radio structures some of which are often unresolved down to
parsec scales \citep{blundell96, blundell98, ulvestad01, lal04, nagar99, nagar02, anderson04, ulvestad05_n4151, singh13},
and only sub-relativistic velocities \citep{middelberg04, ulvestad05_rqq}.  
The physical origin of the radio emission in RQ AGN, whether a downscaled version of the RL, collimated jets \citep{barvainis96, gallimore06}, 
or coronal emission from magnetic activity above the accretion disk \citep{field93}, remains to be resolved.
In some sources, a combination of several spectral components may be present around the nucleus \citep{barvainis96, gallimore04}.  
For the sake of clarity of discussion, in this paper we distinguish a jet that is both highly relativistic and well-collimated, 
from coronal emission that arises from hot gas, and perhaps an outflow that is neither relativistic nor well-collimated.

A connection between the radio and X-ray emission in RQ AGN is
suggested by the correlation of the radio luminosity $L_R$ at 5 GHz and the X-ray luminosity $L_X$  
\citep{brinkmann00, salvato04, wang06, panessa07}.  
\citet{laor08} used the PG quasar sample \citep{schmidt83} to demonstrate that not only are $L_R (\equiv \nu L_\nu$ at 5~GHz) and
$L_{\rm X}$ (0.2 -- 20 keV) correlated over a large range of AGN luminosity, but that
the correlation follows the well established relation for coronally
active cool stars $L_{\rm R}/L_{\rm X}\sim 10^{-5}$ \citep{guedel93}. 
This suggests that radio emission from RQ AGN is due to
magnetic coronal activity, akin to that of stellar coronae.  
Since the $L_R/L_X\sim 10^{-5}$ relation is accepted as a manifestation of
coronal heating by energetic electrons following magnetic reconnection
in cool-stars, which subsequently gives rise to X-ray emission, the
correlation presented in \citet{laor08} over 20 orders of
magnitude in luminosity raises the possibility that radio emission in
RQ AGN may also be related to coronal, magnetic activity.

There are, however, several physical differences between X-ray and radio emission
of stellar coronae and AGN.
X-ray spectra of stellar coronae are thermal with $T \sim 10^6$~K, 
while the non-thermal (Comptonized) spectra of AGN imply $T \sim10^9$~K.  
Moreover, while X-rays from AGN vary dramatically over short time scales, 
the variability at 5~GHz and 8.5 GHz is much slower and smaller in amplitude.  
If radio emission from RQ AGN is due to self absorbed synchrotron, absorption should decrease with frequency.
Consequently, observations at higher frequencies should probe smaller regions of the source,
and  variability time scales of RQ AGN should thus also decrease with frequency. 
Very recently, a theoretical work by \citet[][arXiv 1411.2334]{inoue14} further highlights the potential of mm observations
for detecting coronal magnetic activity on AGN accretion disks.

As opposed to 5 and 8.5 GHz, higher frequency observations of RQ AGN are scarce.
\citet{barvainis96} observed RQ AGN up to 15~GHz.
About half of the RQ AGN in that sample appeared to have flat or inverted spectral components, 
and a hint of variability, strongly suggesting opaque synchrotron emission from the nucleus.
\citet{park13} reach similar conclusions based on observations at 22 and 45 GHz.
Above 300 GHz (sub-mm) there is a steep rise of the spectrum due to the low frequency tail of thermal dust emission \citep{barvainis92, hughes93},
which leaves a largely unexplored band between 50 -- 300 GHz, crudely referred to here as the mm waveband.
Notable exceptions are the works of \citet{doi05, doi11}, who observed low luminosity AGN and early type galaxies.
That sample includes both RQ and RL sources,
some of which feature flat or even inverted spectra at high frequencies, indicating genuine optically thick core emission.  
\citet{owen81}, \citet{owen82}, and \citet{sadler08} studied X-ray selected quasars, but mostly RL AGN brighter than 1 Jy at 90 and 95 GHz, much brighter and more luminous than any of the present sources.

Little data are available on radio monitoring of RQ AGN on short time scales.  
Even less are available for simultaneous radio and X-ray observations. 
Recently, \citet{jones11} published their campaign for NGC\,4051.
Despite the systematic monitoring, the conclusion of that study was
ambiguous due to the limited radio variability at 8.4 and 4.8~GHz. 
It is also worth noting a similar work by \citet{bell11}, but for the RL NGC\,7213.

In this work, we focus on eight RQ AGN that are well known for their
bright X-ray emission and variability.  We observe them at 95 GHz in
order to obtain their high-frequency radio properties.  One aspect we
wish to test is whether they are suitable candidates for
high-frequency radio (and X-ray) monitoring.  Ultimately, we believe
the simultaneous radio and X-ray monitoring of these targets will
enable the exploration of the physical origin of the radio emission in
RQ AGN.  If the aforementioned coronal conjecture is valid, at high
radio frequencies, variability is expected to approach and be closely
related to that of the X-rays, as is often the case in stellar coronal
flares, and manifested, e.g., by the Neupert effect \citep{neupert68}.
The remainder of the paper is organized as follows: In
Sec.~\ref{sec:sample} we describe the observations of the eight objects. In
Sec.~\ref{sec:results} we present the results, including source flux densities
and inferred broad-band radio spectra.  The implications of these
spectra are discussed in Sec.~\ref{sec:discussion} with a comparison
with RL AGN. The conclusions and outlook are given in
Sec.~\ref{sec:conclusions}.

\section{Sample and Observations}
\label{sec:sample}

The current sample was chosen based on the high X-ray brightness and documented variability of its sources.  
This criterion selects high accretion AGN among the RQ ones. 
Although the flux and morphology of most AGN at 95~GHz was essentially unknown, 
we did make sure to choose radio sources that are compact and sufficiently bright ($>$ 1 mJy) at lower frequencies.  
Six of the targets, namely NGC\,3783, NGC\,5548, NGC\,7469, NGC\,3227, Ark\,564, and Mrk\,766,
are highly variable in the X-rays according to the extensive
monitoring of \citet{markowitz04}.  MR\,2251-178 is a bright
X-ray quasar that has been observed often in the X-rays.  The X-ray
light curve of MR\,2251-178 closely correlates with the optical light
curves over a period of 2.5 years
\citep{arevalo08}.  
MR\,2251-178 as well as NGC\,3783 show evidence for variable emission in the near-IR \citep{lira11}.
Finally, NGC\,5506 is the brightest radio source in the sample.
In the X-rays, it reveals X-ray variability on time scales from days to years \citep{uttley05, soldi11}.

We observed continuum emission in the 3 mm (95~GHz) window from all eight targets.  
Three targets were observed with ATCA (the Australia Telescope Compact Array), and  
five targets were observed with CARMA (the Combined Array for Research in Millimeter-wave Astronomy). 
The observation log from
both telescopes is given in Table~\ref{obslog}.

\begin{table*}
 \centering
\caption{Observation log}
\begin{tabular}{l|ccccccc}
  \hline
Object       &  Telescope  & Array & Resolution & Date & Phase  & Pass band & Flux \\
      &   & & (\arcsec) & & calibrator & calibrator & calibrator \\
  \hline
MR~2251-178 &  ATCA   & H214D    & 2.0  & 30-09-2013 & 2243-123 & 3C446 & Uranus \\
NGC~3783    &  ATCA   &   H168D  & 2.0  & 29-03-2014 & 1144-379 & 3C279 & Mars \\
NGC~5506    &  ATCA   &  H168D   & 2.0  & 29-03-2014 & 1406-076 & 3C279 & Mars \\
NGC~7469    &  CARMA  &  C  &  2.2 & 11-11-2013 & 3C454.3 & 1927+739 & mwc349 \\
ARK~564     &  CARMA  &  C  & 2.2  &  16-11-2013 & 2236+284 & 1927+739 & mwc349 \\
NGC~3227    &  CARMA  &  D  & 5.5 & 03-01-2014 & 0956+252 & 3C273 & Mars \\
MRK~766     &  CARMA  &  D  & 5.5 & 19-01-2014 &1159+292  & 3C273 & Mars \\
NGC~5548    &  CARMA  &  D  & 5.5 & 17-02-2014 & 1310+323 & 1635+381 & Mars \\
  \hline
\end{tabular}
\label{obslog}
\end{table*}

ATCA is an array of six 22-m antennas at the Paul Wild Observatory
(Australia). The ATCA data were taken in September 30, 2013 and in March
29, 2014. We used ATCA in an extended array configurations (H214D and H168D)
to provide a good {\it uv} coverage with long baselines (up to 4500 m) and reach an angular resolution of 2\arcsec at 95 GHz.

CARMA is a 15 element interferometer consisting of nine 6.1 meter antennas and six 10.4 meter antennas,
located in California (USA).
The CARMA observations were performed in C- or D-array configuration,
reaching an angular resolution of 2\farcs2 and 5\farcs5, respectively.

The MIRIAD software package \citep{sault95} was used for reduction of
the visibility data, including flagging data affected by shadowed
antennas, poor weather or antenna malfunctions.
Absolute flux, phase, and pass band calibrators are listed in Table  \ref{obslog}.
After obtaining the phase and amplitude solution,
we apply them to the target source using standard procedures.
Eventually, we invert  and clean with natural weight the visibility data to obtain the map. 
The MIRIAD {\it imfit} task is used to measure the source flux density as well as its uncertainty using a point-source fit.



\section{RESULTS}
\label{sec:results}

\subsection{95 GHz Flux Density}
\label{95GHz}

\begin{figure*}
\includegraphics[scale=0.70,angle=0]{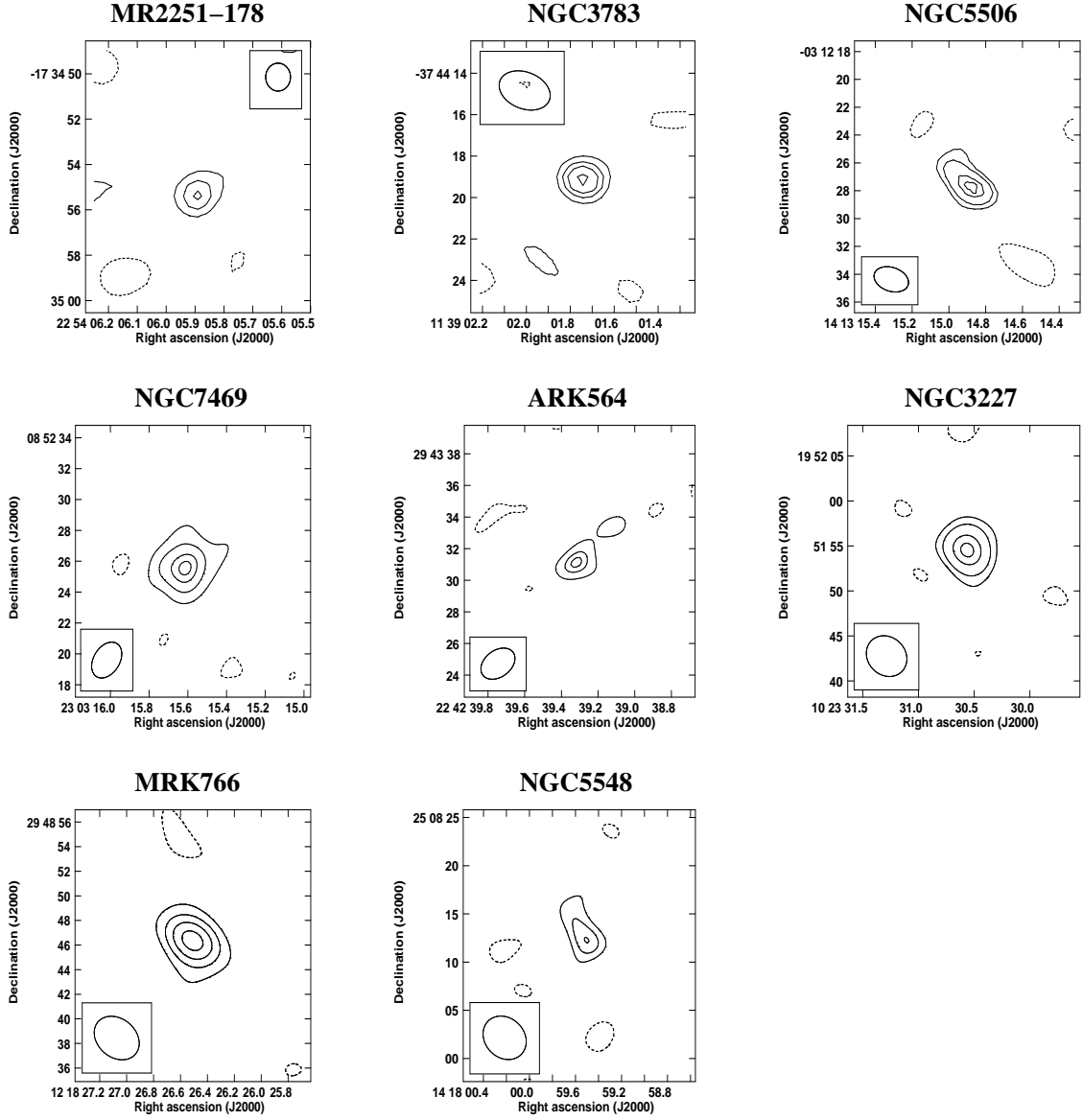} 
\caption{95 GHz contour maps from the present observations. Beam size is indicated for each observation.
Contour levels of the maps are: for MR~2251-178  (-0.4,0.4,0.8,1.2) mJy/beam; for NGC~3783 (-0.8,1.2,1.9,2.3,2.9) mJy/beam; for NGC~5506 (-1.3,2.5,3.8,6.3,7.5) mJy/beam; for NGC7469 (-0.5,1,2,3,4) mJy/beam; for ARK~564 (-0.38,0.54,0.9,1.1) mJy/beam; for NGC~3227 (-0.4,1,1.7,3,3.8) mJy/beam; for MRK~766 (-0.4,0.6,1.0,1.2,1.8) mJy/beam; for NGC~5548 (-0.6,1,1.4,1.6) mJy/beam.}
\label{maps}
\end{figure*}

We obtained radio maps for all of the objects observed
with the CARMA and ATCA telescopes at 95 GHz.  
The maps are presented in Fig.~\ref{maps}.
The measured flux densities range between $\sim$1 to  10 mJy. 
The sources are essentially unresolved, or just resolved at the observed scales (Table~\ref{obslog}).
This indicates that the morphology of the radio source,
whether a narrow relativistic jet, or a broader and slower coronal outflow, can not be assessed
from imaging.  Consequently, the essential physical information of the
source will currently need to come from the broad-band radio spectrum,
and later from variability.

The presently measured flux densities at 95~GHz are listed in the first column of  Table~\ref{radiosphere} along with their uncertainties,
which are generally $<20\%$, and much better for the brightest sources.  
The known distances and consequent luminosities are also listed (with the same relative uncertainties).
These fluxes are likely dominated by core emission.
However, the possible contamination from unresolved radio jets cannot be ruled out.
The physical size in pc of an opaque (self-absorbed) synchrotron source can be estimated from the measured flux
(e.g., Laor \& Behar 2008) to be: 

\begin{equation}
R_{\rm pc} = 0.54L_{30}^{1/2}\nu_{\rm{GHz}}^{-5/4}B^{1/4}
\label{Rpc}
\end{equation}

\noindent 
where $\nu_{\rm{GHz}}$ is the observed frequency in GHz, $L_{30} =
L_{\nu}/10^{30}$ erg\,s$^{-1}$\,Hz$^{-1}$ and $B$ is the magnetic field in Gauss.  
Hence, the size of this optically thick radio-sphere increases with radio luminosity, and decreases with frequency. 
The term radio-sphere is used below when referring to this estimate of the radio-core size based on its luminosity. 
The measured 95~GHz fluxes of the eight sources, therefore, can provide an idea of the sizes of the 95~GHz cores, 
assuming a value $B$ for the magnetic field, although the dependence on $B^{1/4}$ is weak.  
The resulting sizes of the eight present sources are listed in Table~\ref{radiosphere}, assuming $B=1$~G, 
which is of the order expected from equipartition ($B_{\rm eq}^2/8\pi = L_{\rm bol} / 4\pi R_{\rm pc}^2 c$, where $L_{\rm bol}$ is the bolometric luminosity).  
An expression for $R_{\rm pc}$ at equipartition is given in Eq. (22) in \citet{laor08}.
Ignoring the unknown $B$, the (fractional) uncertainty of $R_{\rm pc}$ can be assumed to be half that of the flux density (second column in Table~\ref{radiosphere}), as $R_{\rm pc} \propto L_\nu^{1/2}$, namely $<10\%$.
Evidently, the obtained sizes $R_{\rm pc}$ of the 95~GHz radio-spheres are of the order of $10^{-4}$ to $10^{-3}$ pc.
The light crossing time $R_{\rm pc} / c$ can provide an idea of the minimally expected variability time scales at 95~GHz, namely $0.1 - 1$ day.

The 2--10 keV X-ray luminosities $L_{\rm X}$ (accurate to a few percent) are also listed for comparison in Table~\ref{radiosphere}. 
The 95~GHz to X-ray luminosity ratios of the present sample are rather tightly concentrated around $L_R/ L_X = 10^{-4}$,
with $\log (L_R/ L_X) = -4.1\pm0.25$ (Table~\ref{radiosphere}).
The chance probability for such a tight relation is only $<0.00075$. 
How this compares with other samples is discussed in Sec.~\ref{sec:RX}.

\begin{table*}
\begin{center}
\caption{Luminosity and radio-sphere size properties}
\begin{tabular}{lccccccc}
\hline
Object       & $F_{\rm 95\,GHz}$ & $D_{\rm L}$  &  $L_{\rm 95\,GHz}$   & $R_{\rm pc}$ & $\log L_{\rm R}$(95\,GHz)& $\log L_{\rm X}$ & $\log(L_R/L_X)$ \\[0.1cm]
& mJy & Mpc  &  10$^{30}$erg\,s$^{-1}$Hz$^{-1}$  & $10^{-3}$\,pc & erg\,s$^{-1}$ & erg\,s$^{-1}$ \\
\hline
MR\,2251-178 & 1.6$\pm$0.3 & 289.1& 0.16$\pm$0.03 &  0.70 & 40.18  &   44.40 & --4.22    \\
NGC\,3783  & 3.1$\pm$0.7  & 42.2  & 0.0066$\pm$0.0015 &  0.15 & 38.78  &   43.15 &--4.35    \\
NGC\,5506   &  10$\pm$1.0 & 26.8  & 0.009$\pm$0.001 &  0.17 & 38.91 &   42.74 & --3.83  \\
NGC\,7469    & 4.95$\pm$0.16 & 71.2  & 0.030$\pm$0.001  &  0.32 & 39.46  &   43.18 &--3.72 \\
ARK\,564    & 1.14$\pm$0.19 & 108.4 & 0.016$\pm$0.003 &  0.23 & 39.18  &   43.57 & --4.39   \\
NGC\,3227    &  4.1$\pm$0.24 & 16.7  & 0.0014$\pm$0.0001  &  0.07 & 38.12  &   41.90 &--3.78  \\
MRK\,766     &  1.98$\pm$0.17  &  56.3 & 0.0075$\pm$0.0006 &  0.16 &  39.10   &   43.28 & --4.18  \\
NGC\,5548    & 1.6$\pm$0.3 & 75.0  & 0.011$\pm$0.002  &  0.19 &  38.76  &   42.93 &--4.17  \\
\hline
\end{tabular}
\label{radiosphere}
\end{center}
\end{table*}

The black hole mass $M_{\rm BH}$, $R_{\rm pc}$ in terms of gravitational radii $r_g = GM_{\rm BH} / c^2$, and
$L_{\rm bol}/L_{\rm {Edd}}$ are listed in Table~\ref{MBH}.
Evidently, the radio-sphere sizes correspond to $\approx 10 - 1000~r_g$.
The bolometric to Eddington luminosity ratio $L_{\rm bol}/L_{\rm {Edd}}$ of the present sample also spans more than two orders of magnitude from 0.01 -- 3.0. These ranges are much larger than the uncertainties in $M_{\rm BH} (L_{\rm {Edd}})$ and $L_{\rm bol}$, 
which are dominated by systematics, and can reach a factor of 3 \citep{peterson04}. 
In the present small sample, we confirm that $L_{\rm bol}$ scales with $L_{\rm 95GHz}$,
which is consistent with low-frequency results \citep[$L_{\rm bol} \propto L_\nu ^{1.2}$ in][]{white07} 
and with radio galaxies \citep[$L_{\rm bol} \propto L_\nu$ in][]{sikora13}.

The factor 10 range in $R_{\rm pc} (\propto L_\nu^{1/2}$, Table~\ref{radiosphere}),  and factor of 100 range in $r_g (\propto M_{\rm BH}$, Table~\ref{MBH}) results in
an inverse correlation of $R_{\rm pc} /r_g$ with $M_{\rm BH}$.
The four low-mass black holes with $M_{\rm BH} \le 10^7$ all have $R_{\rm pc} /r_g \ge 330$, and the four massive ones with $M_{\rm BH} \ge 3\times 10^7$ all have $R_{\rm pc} /r_g \le 52$.
Interestingly, $R_{\rm pc} / r_g$ appears to increase with $L_{\rm bol}/L_{\rm {Edd}}$ in the sense that the four 
low $L_{\rm bol}/L_{\rm {Edd}} (< 0.1)$ sources have small (weak) radio-spheres of a few $10 r_g$,
while the four high $L_{\rm bol} / L_{\rm {Edd}} (>  0.4)$ sources have large (luminous) radio-spheres of $300 - 1000~r_g$, implying that higher accretion rates ($L_{\rm bol}/L_{\rm {Edd}}$) are associated with mm-wave activity on larger relative scales.

\begin{table*}
\begin{center}
\caption{Black hole mass, size of radio-sphere in gravitational radii, and bolometric luminosity}
\begin{tabular}{lcccccc}
\hline
Object       & $M_{\rm BH}$ & Ref ($M_{\rm BH}$) & $R_{\rm pc}$ & $\log L_{\rm bol}$ & Ref ($L_{\rm bol}$) & $L_{\rm bol}/L_{\rm {Edd}}$  \\[0.1cm]
& 10$^6$M$_{\odot}$ & & $r_g$ & erg\,s$^{-1}$ & & \\
\hline
MR\,2251-178 & 200 & \citet{lira11} & 38 & 45.3 &  \citet{lira11} & 0.08  \\
NGC\,3783  & 29.8 & \citet{peterson04} & 52 & 44.4 & \citet{woo02} & 0.07   \\
NGC\,5506   &  2.5 & \citet{nagar02b} & 710 & 44.1 & \citet{soldi11} & 0.4 \\
NGC\,7469    & 10 & \citet{peterson14} & 330 & 45.3 & \citet{woo02} & 1.5  \\
ARK\,564    & 2.6 & \citet{botte04} & 930 & 45.0 & \citet{romano04} & 3.0 \\
NGC\,3227    &  42.2 & \citet{peterson04} & 17 & 43.9 & \citet{woo02} & 0.014 \\
MRK\,766     & 1.8 & \citet{bentz09} & 920 & 44.0 & \citet{bentz09} & 0.4  \\
NGC\,5548    &  67.1 & \citet{peterson04} & 30 & 44.8 & \citet{woo02} & 0.08  \\
\hline
\end{tabular}
\label{MBH}
\begin{flushleft}
$M_{\rm BH}$ is taken directly from the literature, except for NGC\,5506 in which it is estimated by using the broad Paschen line width and luminosity from \citet{nagar02b}.
\end{flushleft}
\end{center}
\end{table*}

\subsection{95 GHz Excess}
\label{sec:excess}

We used the literature to obtain the radio flux density of the present sources at frequencies below 95~GHz.  
We found measurements from 1.4~GHz up to the highest radio frequency available of 22 GHz for NGC\,5506 and NGC\,7469.  
When the core is resolved in the archival data, we only consider that component to probe the emission closest around the nucleus.  
In order to obtain the highest spatial resolution, we first consider radio detections from the VLA in the high-resolution A-array configuration.
Table~\ref{coreflux} lists these archival core flux density measurements, as well as the current 95~GHz measurement.
Since the current measurements have larger beams than those of the VLA A-array in the different bands,
we also present in Table~\ref{fluxlargebeam} archival flux density measurements obtained with other VLA configurations which basically have larger beams than in Table~\ref{coreflux}.

\begin{table*}
\begin{center}
\caption{Array-A VLA flux densities in mJy. Excess factor at 95~GHz is relative to the low-frequency spectrum.}
\begin{tabular}{l|cc|cc|cc|cc|cc|cc}
\hline
 Frequency    &   \multicolumn{2}{c|}{1.4 GHz}   &  \multicolumn{2}{c|}{5 GHz}  &  \multicolumn{2}{c|}{8.5 GHz}  &  \multicolumn{2}{c|}{15 GHz}  &  \multicolumn{2}{c|}{22 GHz} &   \multicolumn{2}{c|}{95 GHz}  \\
Resolution   &\multicolumn{2}{c|}{0\farcs3} & \multicolumn{2}{c|}{0\farcs33}& \multicolumn{2}{c|}{0\farcs20} &  \multicolumn{2}{c|}{0\farcs13}  & \multicolumn{2}{c|}{0\farcs089} & \multicolumn{2}{c|}{2\farcs2-5\farcs5} \\
\hline
 Object            &   $F_\nu$          &   Ref           &  $F_\nu$        &        Ref  &   $F_\nu$  &          Ref   &     $F_\nu$        &  Ref       &   $F_\nu$       &    Ref  & $F_\nu$  & Excess \\
\hline
MR\,2251-178 &     5.0$\pm$0.1 & map  & 3.1$\pm$0.1 & map   &      ---  &  --- &    ---    &  ---&    --- &  ---  &  1.6$\pm$0.3 & 1.7--27.7 \\
NGC\,3783    &    43.6$\pm$2.0 & NVSS   & 13.0$\pm$1.0  & UL84    &  7.7$\pm$0.2 &  SC01     &    ---  & ---   &  ---       &---  & 3.1$\pm$0.7 & 4.1 \\
NGC\,5506    &    304$\pm$5.0  & OR10     & 136$\pm$4.0 & OR10    & 84$\pm$4  & OR10     & 44$\pm$5.0  & UL84     &  22$\pm$4  & TA11   &  10$\pm$1 & 1.2-1.3  \\
NGC\,7469    &    134$\pm$6.0  & UN87 & 26.9$\pm$2.0 & LA04  &  14.8$\pm$0.3 & OR10    & 10.6$\pm$0.3  & OR10  &  17.5$\pm$0.5  & PR04  & 5.0$\pm$0.2 & 1.3--1.7 \\
ARK\,564     &    25.7$\pm$0.2 & map   & 8.6$\pm$0.4 & LA11    & 3.0$\pm$0.5  & SC01      &     ---        & ---  &   ---  &  ---  &  1.1$\pm$0.2  & 4.8 \\
NGC\,3227    &    78.2$\pm$5.0   & HO01     &  15$\pm$2   & UL81   &  12.2$\pm$1.3 &  KU95  &  4.7$\pm$0.3  & NA05   & ---    &   --- &  4.1$\pm$0.2  & 5.3-6.8 \\
MRK\,766     &    36.4$\pm$1.9 & UL89     & 11.7$\pm$0.6 & UL95  &  8.7$\pm$0.4 & KU95 &  ---    & ---   &      ---   & --- &  2.0$\pm$0.2   &  1.0-1.8\\
NGC\,5548    &    5.1$\pm$0.5 & HO01    &   4.2$\pm$0.4 & WR00  & 3.9$\pm$1.8 & WR00     &    ---      &     ---    & --- &  ---& 1.6$\pm$0.3 & 0.6-2.8 \\
\hline
\end{tabular}
\label{coreflux}
\begin{flushleft}
Reference (beam size indicated in parenthesis when deviates from nominal): NVSS (45\arcsec)  \citet{NVSS}, UL84
  \citet{ulvestad84a}, SC01 \citet{schmitt01}, OR10 \citet{orienti10},
  TA11 \citet{tarchi11}, UN87 \citet{unger87}, LA04  (1\arcsec) \citet{lal04},
  LA11 \citet{lal11}, UL81 \citet{ulvestad81}, KU95 \citet{kukula95}, PR04 (2\farcs5) \citet{prouton04},
  NA05 \citet{nagar05}, HO01 \citet{ho01c}, UL89 \citet{ulvestad89},
  and WR00 \citet{wrobel00}. 
  'map' stands for the map obtained from the VLA archive; For MR\,2251-178 map beam sizes are 6\farcs6 at 1.4\,GHz and 1\farcs8 at 5\,GHz.
\end{flushleft}
\end{center}
\end{table*}

\begin{table*}
\begin{center}
\caption{Flux densities in mJy and beam size in arcseconds from VLA observations with larger beam sizes than in Table~\ref{coreflux}.}
\begin{tabular}{l|cc|cc|cc|cc|}
\hline
 Frequency:     &   \multicolumn{2}{c|}{1.4GHz}   &  \multicolumn{2}{c|}{5 GHz}  &  \multicolumn{2}{c|}{8.5 GHz}  &  \multicolumn{2}{c|}{15 GHz}  \\ 
\hline
 Object            &   $F_\nu$          &   Ref  (beam)    &  $F_\nu$        &   Ref  (beam) &   $F_\nu$  &   Ref  (beam)  &     $F_\nu$        &  Ref  (beam)   \\
\hline
MR\,2251-178 &    16.2$\pm$0.1 & NVSS (45\arcsec)   & ---    &   ---   &  --- &    ---    &  ---&    ---   \\
NGC\,3783 &    --- & ---   & ---    &   ---   &  --- &    ---    &  ---&    ---   \\
NGC\,5506    &    337.9$\pm$0.1  & FIRST (5\arcsec) & 139.2$\pm$0.1 & CO96 (3\arcsec)   & ---  &    ---  & 58$\pm$3.0  & OR10 (0\farcs7)  \\
NGC\,7469    &    145.7$\pm$0.1  & FIRST (5\arcsec) & 66.1$\pm$3.3 & ED87 (15\arcsec)   &  42$\pm$4.0 & KU95 (2\farcs5)   & ---  & ---    \\
ARK\,564     &    28.6$\pm$0.2 & NVSS (45\arcsec)   & 10.7$\pm$0.2 & LA04 (1\arcsec)   &  ---  & ---     &     ---        & ---  \\
NGC\,3227    &    82.8$\pm$0.1  & FIRST (5\arcsec) &  35.0$\pm$0.7  & GA06 (15\arcsec)   &  16.8$\pm$0.5 &  KU95 (2\farcs5)   &  4.44$\pm$0.6  & map (0\farcs8)  \\
MRK\,766     &    40.3$\pm$0.2 &  FIRST (5\arcsec) & 20.4$\pm$0.6 & GA06 (15\arcsec)   &  10.3$\pm$0.4 & KU95 (2\farcs5) &  ---    & ---    \\
NGC\,5548    &    24.4$\pm$0.2 & FIRST (5\arcsec)  &  10.5$\pm$0.5 & ED87 (15\arcsec) & 4.4$\pm$0.3 & KU95 (2\farcs5)    &    ---      &     ---    \\
\hline
\end{tabular}
\label{fluxlargebeam}
\begin{flushleft}
  Reference: NVSS \citet{NVSS}, FIRST \citet{becker95}, CO96
  \citet{colbert96}, OR10 \citet{orienti10}, ED87 \citet{edelson87},
  KU95 \citet{kukula95}, LA04 \citet{lal04}, and GA06 \citet{gallimore06}. 'map' stands for the map obtained
  from the VLA archive. 
\end{flushleft}
\end{center}
\end{table*}

\begin{figure*}
\begin{center}
\includegraphics[scale=0.30,angle=90]{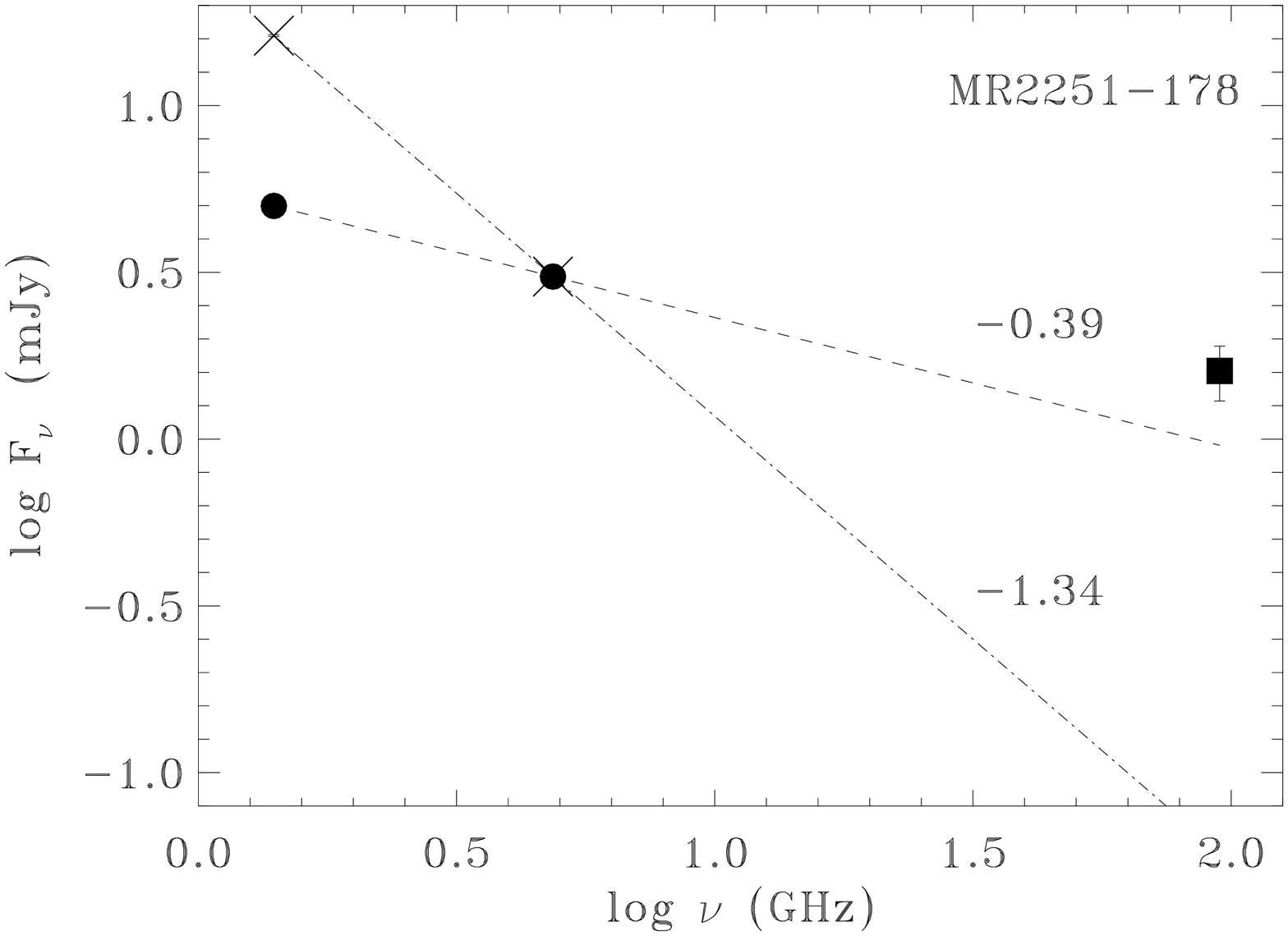} 
\includegraphics[scale=0.30,angle=90]{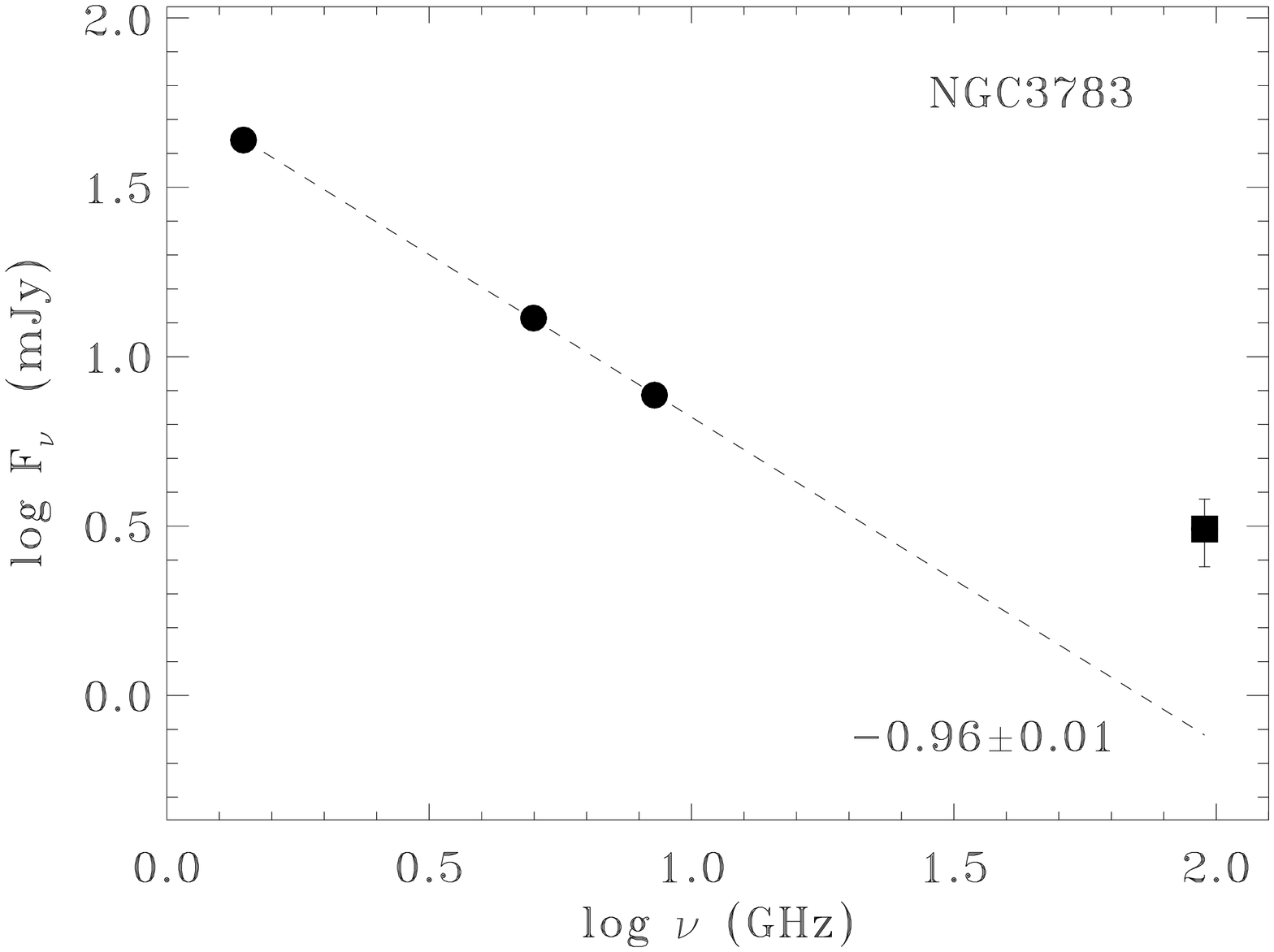} 
\includegraphics[scale=0.30,angle=90]{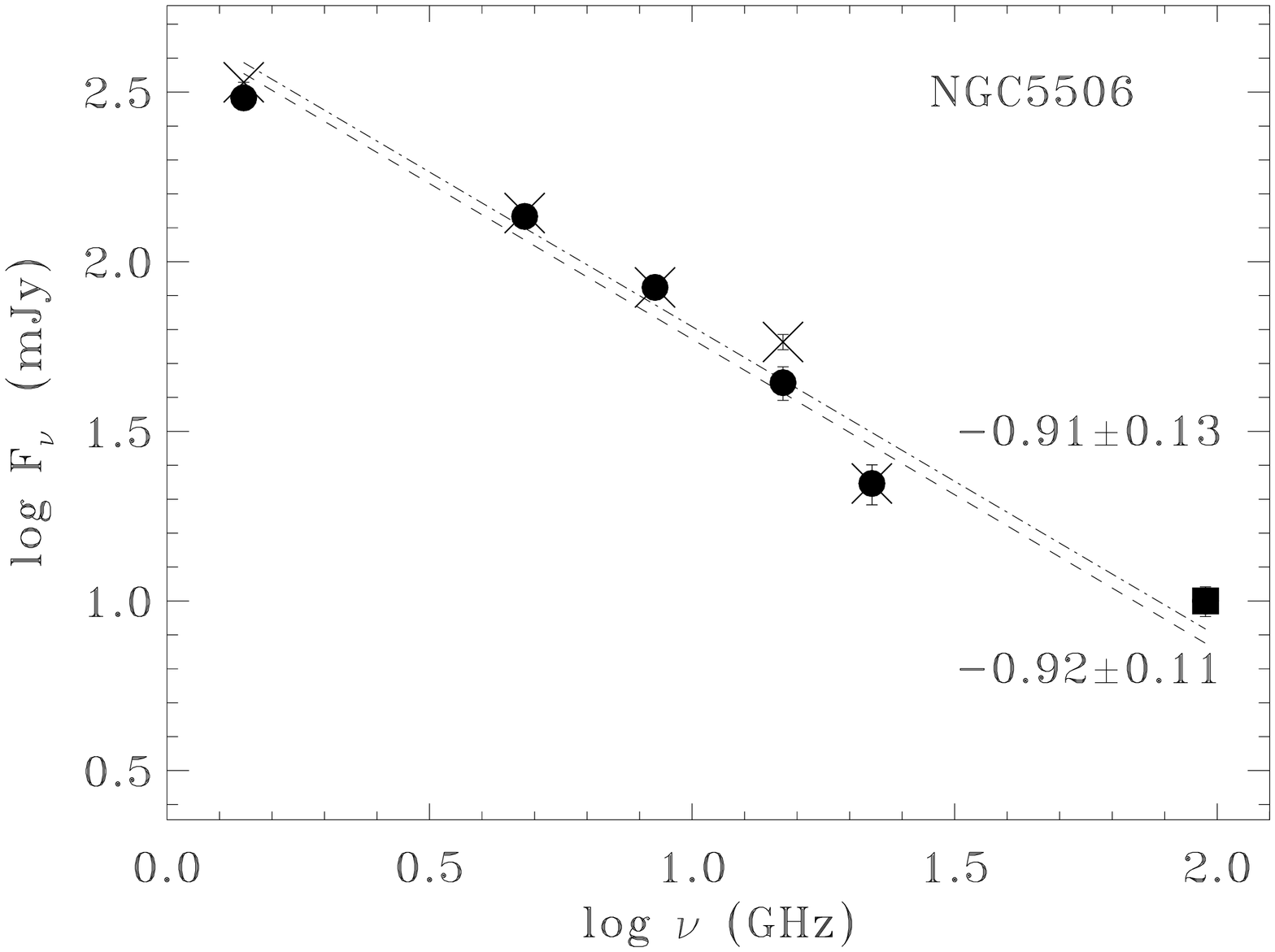} 
\includegraphics[scale=0.30,angle=90]{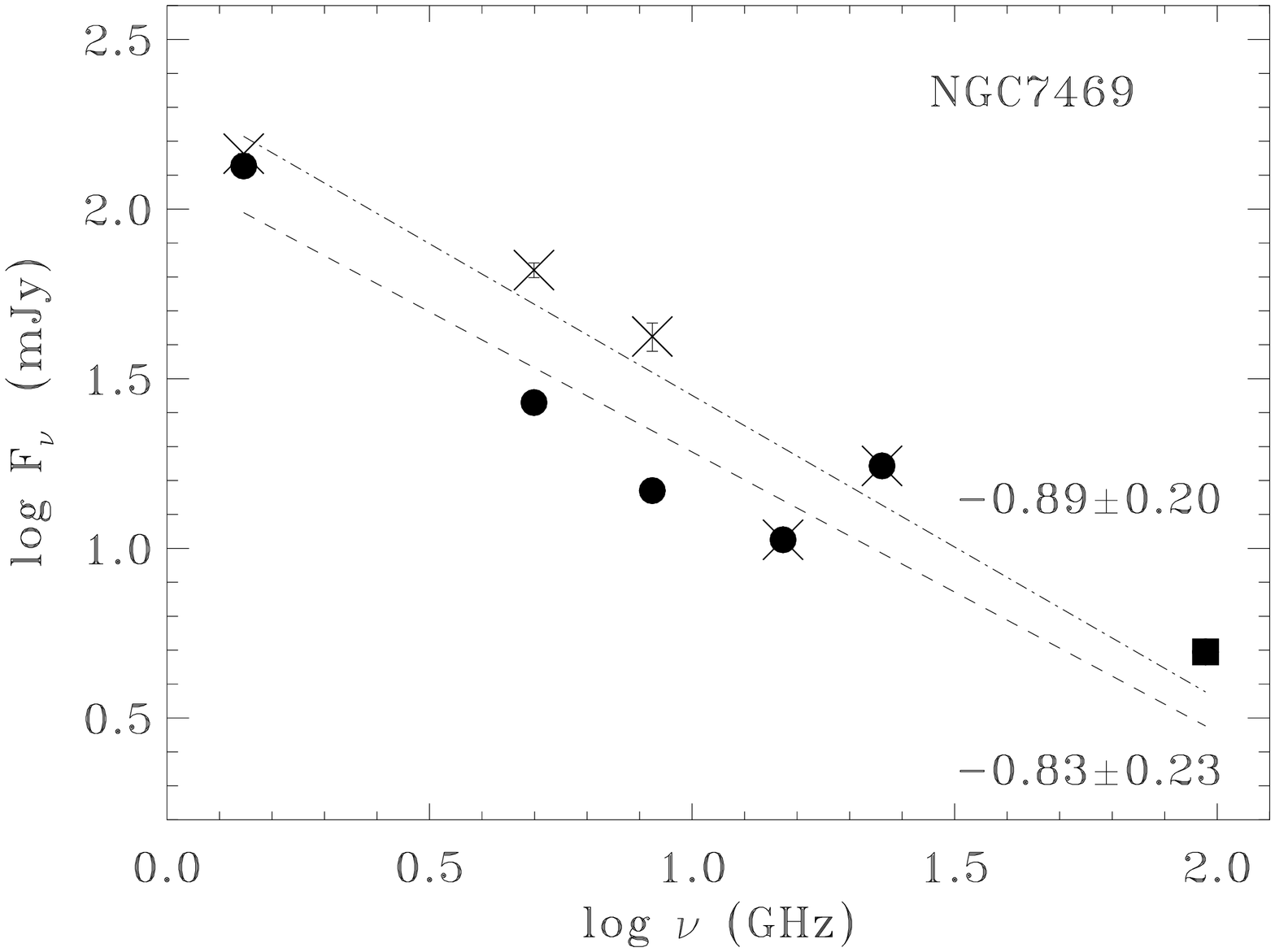} 
\includegraphics[scale=0.30,angle=90]{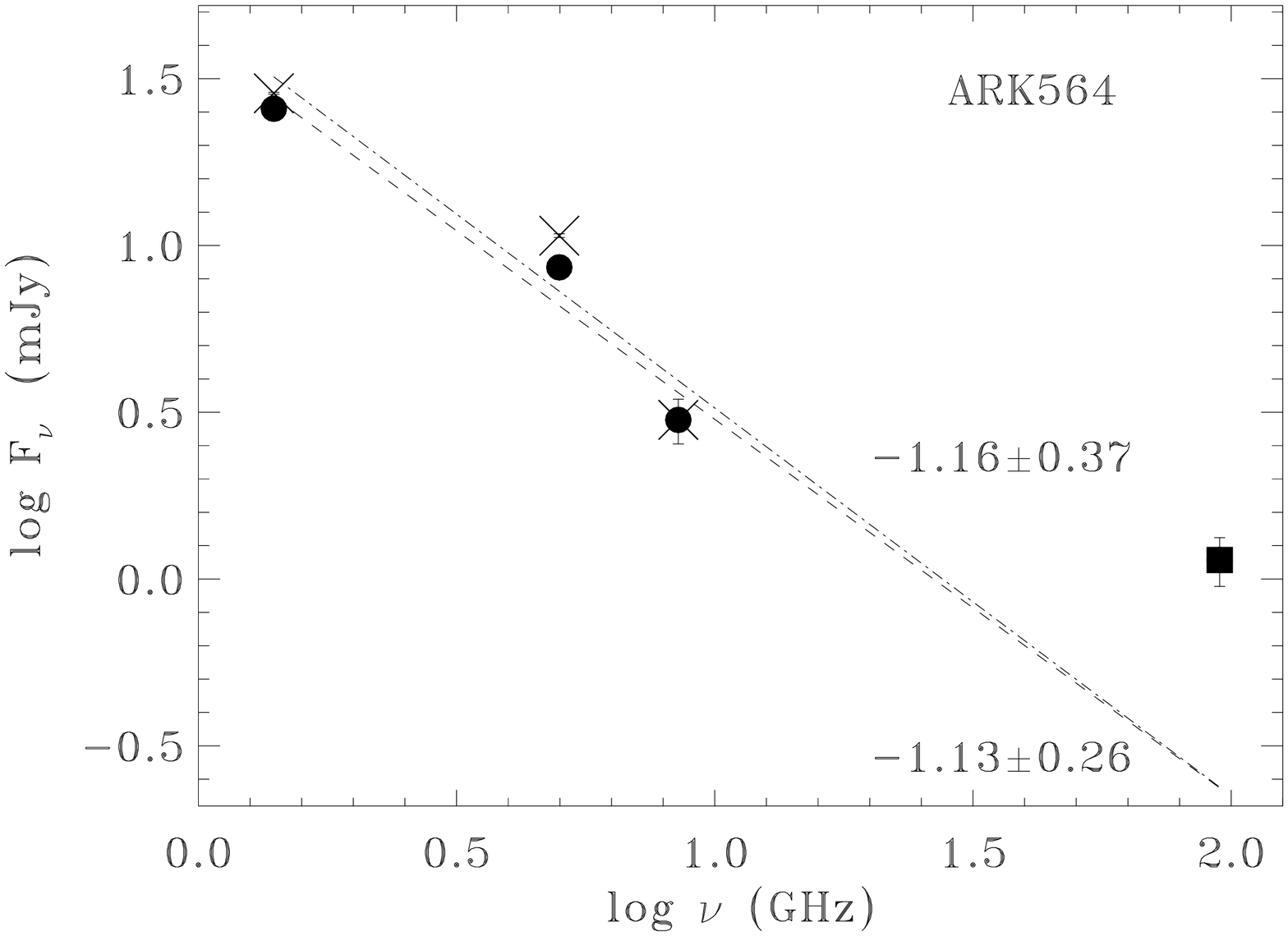}  
\includegraphics[scale=0.30,angle=90]{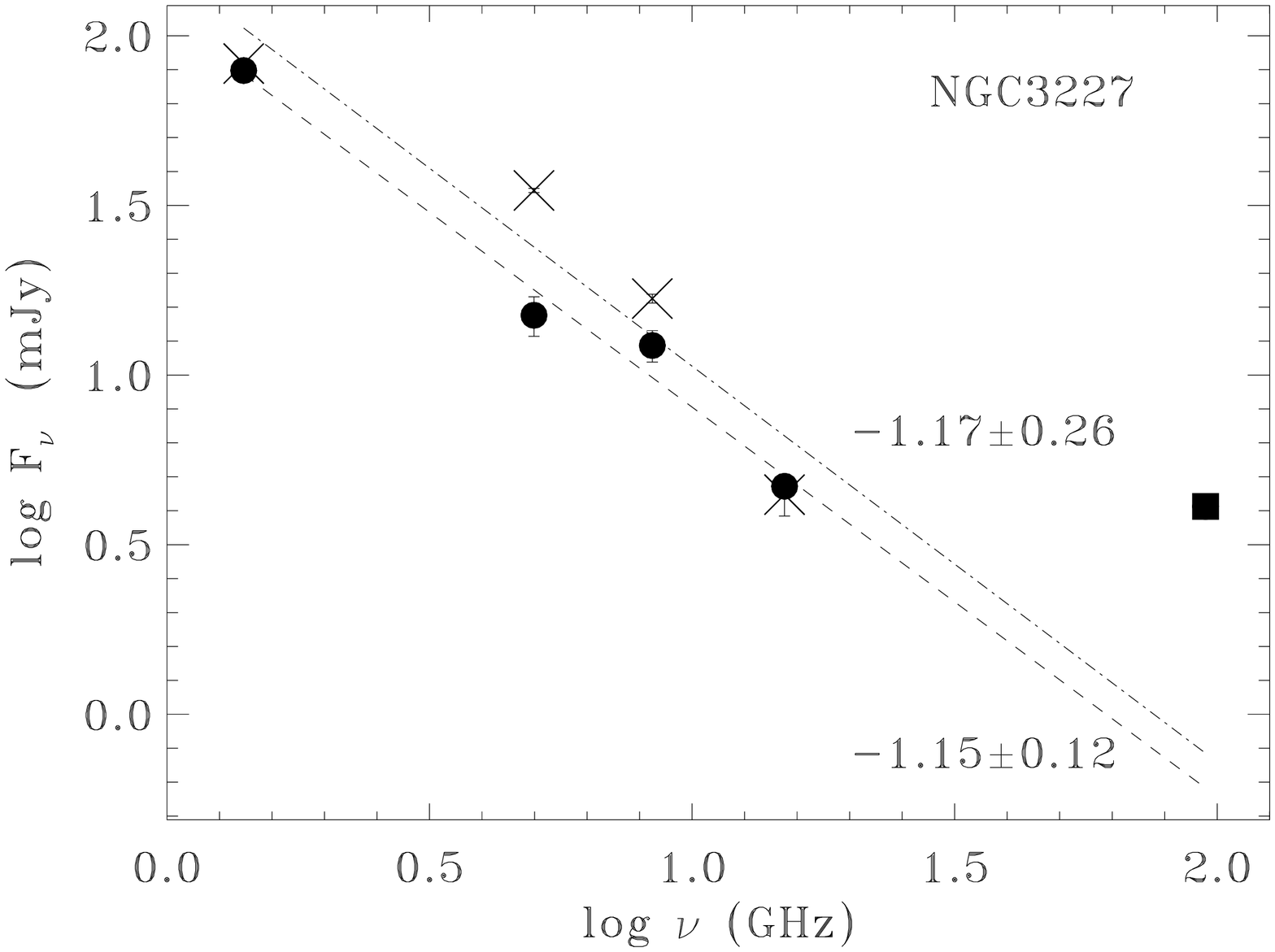} 
\includegraphics[scale=0.30,angle=90]{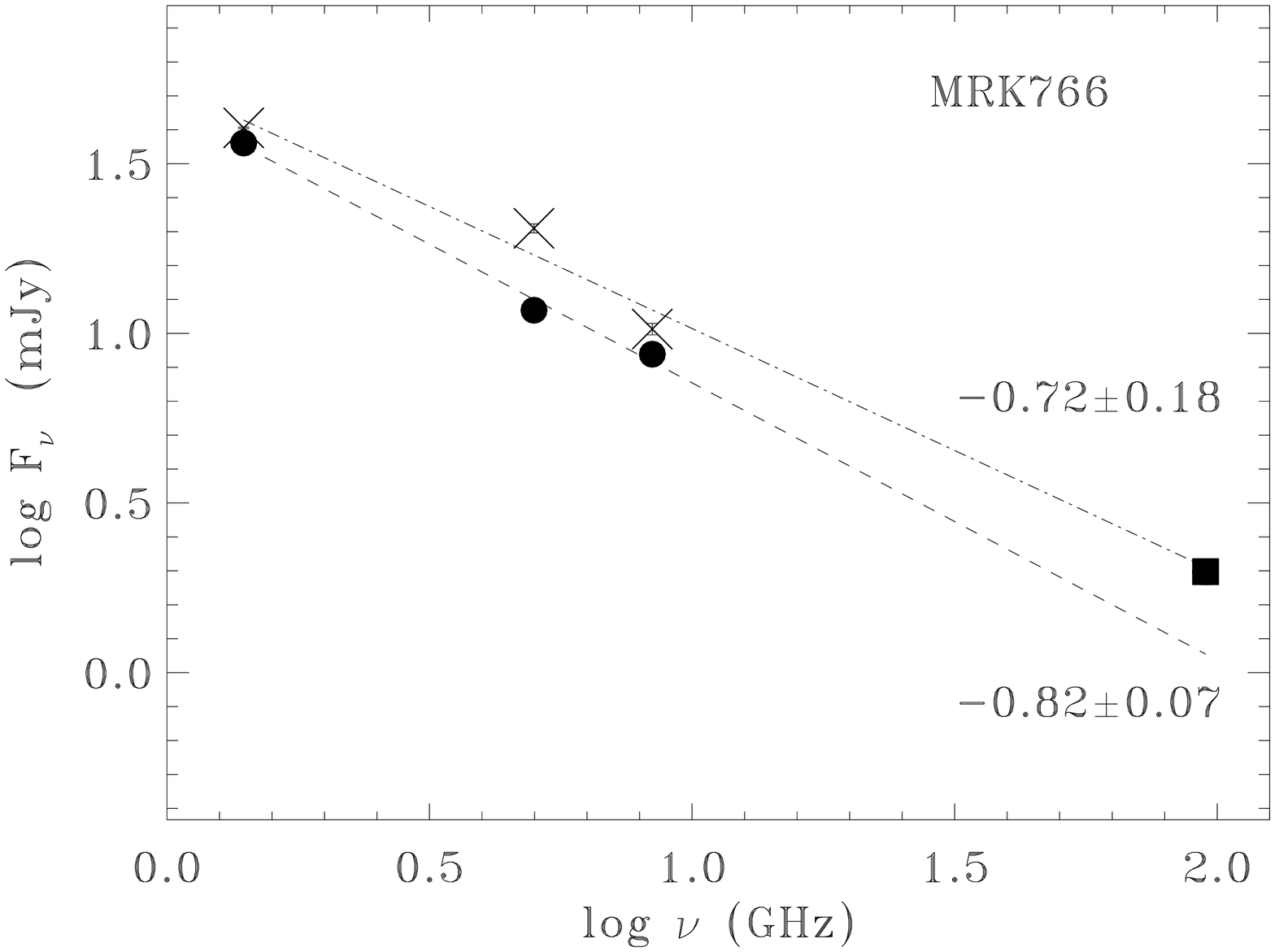} 
\includegraphics[scale=0.30,angle=90]{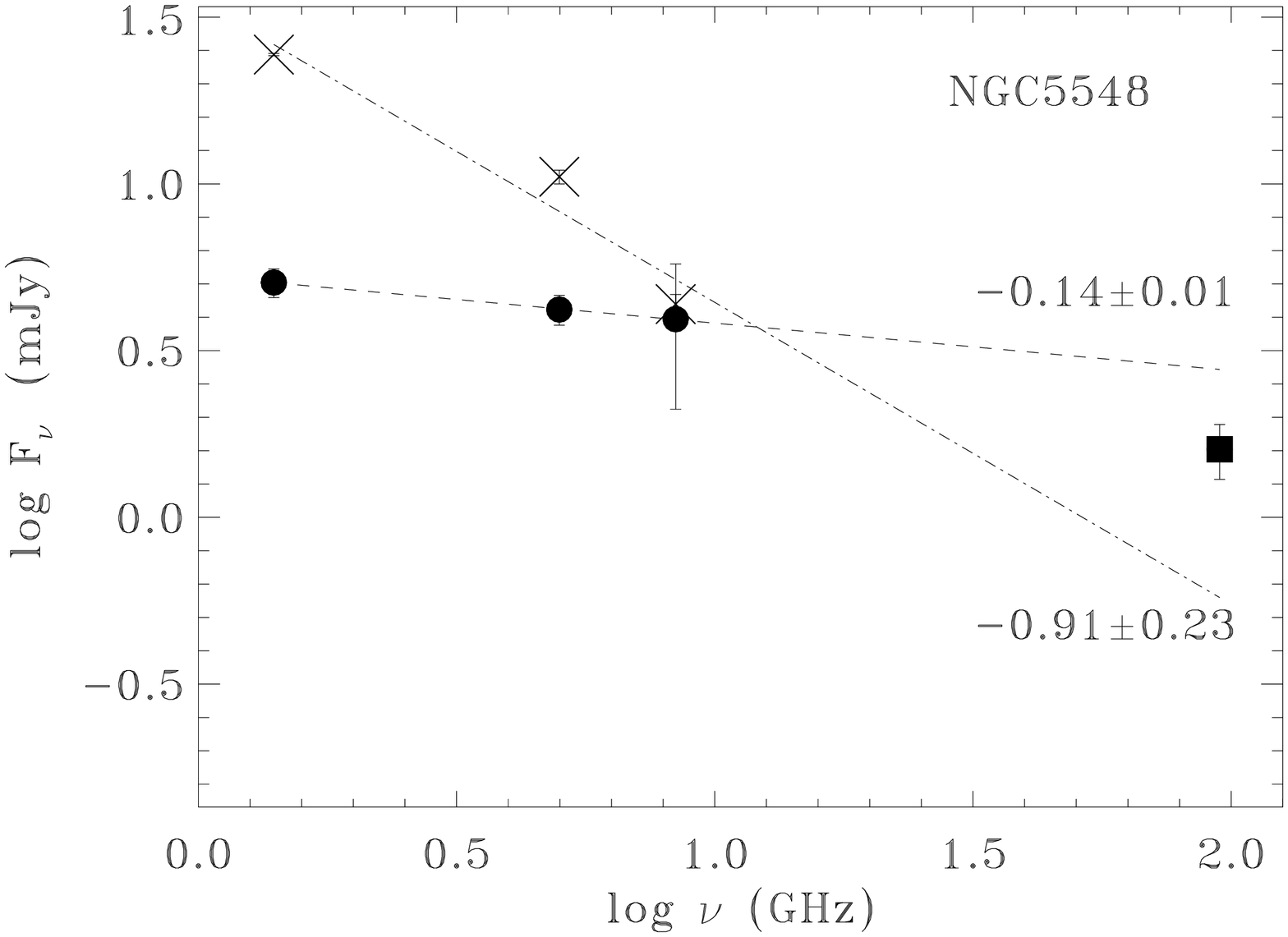}
\end{center}
\caption{
Broad band radio spectral distributions of the eight detected AGNs. 
95 GHz flux densities (squares) are from the present survey, while lower frequency measurements (circles) are taken from the VLA, as much as available.
  Note that configuration and beam size vary dramatically between observations.
  The dashed line is the slope (value indicated) fitted to the high-resolution A-array flux densities (circles).
  Where available, a dashed-dotted slope is provided for flux densities obtained with larger beam size ($\times$).
 In both cases, the 95~GHz measurement is excluded from the fit in order to give an idea of the 95~GHz excess.
Error bars represent measurement uncertainties, except for NGC\,5548 where they reflect variability.}
\label{sed}
\end{figure*}

The derived radio broadband Spectral Energy Distributions (SEDs) are
plotted separately for each source in Figure~\ref{sed}, including also the present detection at 95 GHz. 
One must bear in mind that the unmatched beam sizes at different frequencies may lead to inconsistent radio spectra
if different scales of radio emission are present, such as an extended component from a putative jet.
However, large flux discrepancies between different beams that have an appreciable effect on the spectrum are found only for MR\,2251-178 and for NGC\,5548 (see Figure~\ref{sed}), indicating the flux of the other sources likely originates in the inner sub-arcsec region of the AGN.
Moreover, since the data in Table~\ref{coreflux} are taken over many years, 
and these sources have been observed to vary, the SEDs in Figure~\ref{sed} should be treated with caution.
Below we describe the observed spectral pattern found for these sources.

In Fig.~\ref{sed}, we show the spectral slope indices $\alpha$ ($F_{\nu} \propto \nu^{-\alpha}$),  
obtained from the low-frequency flux densities, excluding the 95~GHz data point.  
When available, we provide one slope for the VLA A-array flux density (circles), and one for the larger beams (triangles).
The spectra for the most part are steep with $0.8 < \alpha < 1.3$.
Mrk\,766 has a somewhat flatter spectrum with $\alpha = 0.68$ when the large beam results are considered.
MR\,2251-178 and NGC\,5548 have flatter spectra ($\alpha = 0.39, 0.14$, respectively),
 when the NVSS (45\arcsec) flux density at 1.4~GHz is ignored.
For the most part, the 95 GHz flux density overshoots the low-frequency slope.
The excess factor for each source is given at the rightmost column of Table~\ref{coreflux}, 
providing a range when two slopes with different beam sizes are available.
The excess is large ($\times 4 - 7$) for NGC\,3783, ARK\,564, and NGC\,3227, and smaller (but still $> 1$) for MR\,2251-178, NGC\,5506, and NGC\,7469. For Mrk\,766 and NGC\,5548 the picture is murkier, although some excess can not be ruled out.
Apart from the mere detection, this 95~GHz excess is the main result of the present campaign.

It is difficult from our data to pinpoint the frequency at which the spectrum turns over from steep to flat or inverted. 
\citet{antonucci88} and \citet{barvainis96} identified spectra that turn over already at 10-- 15~GHz.
They too ascribed this to a self absorbed synchrotron component, and termed it the High-Frequency Excess.
In the present sample, the sources that have archival data above 10~GHz are only NGC\,5506, NGC\,7469, and NGC\,3227.
Only NGC\,7469 shows a possible turnover below the 95 GHz data point (see Fig.~\ref{sed}).
In contrast with those previous works, in which some sources had a high-frequency excess and some did not,
the present sources may all have at least some 95~GHz excess with respect to the lower-frequency slope.

\subsection{Beam sizes and variability}
There are caveats to these results, however.
Since VLA beam sizes comparable to those at 95~GHz (2\farcs2 --5\farcs5 here) are not available at all frequencies,
and in particular since the 15~GHz and 22~GHz observations are at sub-arcsec resolution,
the magnitude of the 95 GHz excess reported above would change if there is extended emission on scales of a few arcseconds, 
but not if the 95 GHz flux is genuine core emission, as suggested by the radio-sphere sizes in Table~\ref{radiosphere}.
Moreover, since the observations are often many years apart, 
the 95 GHz excess needs to be confirmed with simultaneous multi-frequency  observations.

Radio variability of RQ AGN has been well documented \citep{wrobel00, barvainis05, anderson05,  mundell09, jones11}.
\citet{wrobel00} monitored NGC\,5548 at 5 and 8.5 GHz and found variability of up to $\sim$52\% on time scales of months to years,
which is reflected in the larger error bars for this source in Fig.~\ref{sed},
and which is much larger than individual measurement uncertainties (plotted for the other sources).
We further note that NGC\,5548 is in a historically low X-ray and UV flux state for the past few years \citep{Kaastra14}.
If the 95~GHz emission is coming from similar inner regions as  the X-ray and UV, 
it is not surprising that the 95~GHz flux now is much lower than expected.
Given that all of the low-frequency measurements of NGC\,5548 are from the 1990s (see references in Table~\ref{coreflux}),
we suspect NGC\,5548 would feature a stronger 95~GHz excess if all its radio measurements were simultaneous.
In summary, it would be useful to confirm the 95 GHz excess for the entire present sample with a contemporary multi-frequency campaign.


\section{Discussion}
\label{sec:discussion}

Since none of our sources is resolved, we need to base our analysis of the core emission on the spectrum.
On one hand, a steep SED with $\alpha \ga 0.5$ is typical of optically-thin synchrotron emission, 
while a flat SED ($\alpha < 0.3$) is interpreted as the superposition of several optically-thick regions, 
where the synchrotron emission is self-absorbed.  
Steep spectra have been customarily associated with extended jet emission on a kpc scale, 
while flat spectra likely come from a pc-scale core.
These behaviors have been commonly used to interpret both RL and RQ AGN emission \citep[e.g.,][]{barvainis96},
although the power and speed of the jets, if they can be called jets, 
in RQ AGN is believed to be sub-relativistic \citep[e.g.,][]{falcke96, ulvestad98, ulvestad05_rqq}.  
Recently, \citet{kharb15} used parsec-scale VLBI resolution to suggest that these steep-spectrum jets in RQ AGN are in fact coronal winds.
The low-frequency radio emission from RQ AGN is a mixed bag of emission mechanisms,
including complex morphology and varying spectra \citep{barvainis96, ulvestad98, gallimore06, giroletti09}.
Here, we discuss the possibility, based on the small $R_{\rm pc}$ sizes that the 95~GHz cores are due to the accretion disk or to
outflows in its vicinity that are not jetted in nature, namely are neither relativistic nor well collimated.
These could be similar to stellar coronal mass ejections \citep[CMEs; e.g.,][]{bastian98, bastian07}, 
but on the larger scales of an AGN accretion disk.



A similar effect of 95~GHz excess emission (diminishing steepness with frequency, and in some cases even an inverted spectrum) was observed by \citet{doi05, doi11} and by \citet{park13} in low luminosity AGN as well as in early type galaxies. 
Indeed, \citet{doi11} conclude that this is indicative of core AGN emission.  
Extended stellar dust emission is also discussed, but is unlikely, as dust
emission from stars, let alone the AGN, is hotter and becomes important only at higher frequencies above 300~GHz.  
Moreover, \citet{doi11} report variability of several sources on short time scales, which rules out a stellar component.
Interestingly, the most conspicuous 95~GHz excess in the sample of \citet{doi11} 
is found in sources that also reveal very broad optical emission lines when observed at high angular resolution,
namely in NGC\,2782, NGC\,4143, and NGC\,4579 \citep{shields07}.

The mm-wave emission is hardly absorbed once it leaves the radio core,
except perhaps by a Compton thick column of ionized gas, and since free-free absorption strongly decreases with frequency.
Consequently, we expect that the 95~GHz excess will be observed in type II AGN as well.
In fact, mm-wave surveys could exploit the excess to efficiently detect obscured AGN.
Indeed, \citet{gallimore04} find an inverted spectrum for the radio spot (S1) that is believed to be the location of the nucleus of the archetypical type II AGN, NGC\,1068.


In an attempt to explain this spectral change around 100~GHz, \citet{doi11} invoke advection dominated accretion flows (ADAF), since similar behavior is observed in Sgr A*.
However, the present X-ray selected sample is quite different from that of \citet{doi11}, who chose their targets based on their low luminosity.
The present sample includes high $L_{\rm bol}/L_{\rm {Edd}}$ sources, as listed in Table~\ref{MBH} (exceeding unity for NGC\,7469 and Ark\,564).
Therefore, their accretion disks are highly radiative \citep{shakura73}, and ADAF is irrelevant.

We thus suggest the observed high-frequency excess emission is of coronal nature, in the sense that magnetic activity around the accretion disk is responsible for the mm wavelength component. It originates further in the AGN core 
where the energy density of the electrons is higher and the magnetic field is substantially stronger
compared to the more extended regions, from where the low frequency optically-thin synchrotron component arises.
The magnetic activity of the disk may increase with accretion ($L_{\rm bol}/L_{\rm {Edd}}$), and results in increased 95~GHz emission.
In the next subsection, we further examine the place of our sample in the context of other 95~GHz measurements from the literature.

\subsection{Radio to X-Ray Luminosity Ratio}
\label{sec:RX}

In order to further study the origin of the 95~GHz emission of RQ AGN and their possible connection with the accretion disk, we wish to compare with the X-ray emission, which is known to arise from the center of the AGN and to be unbeamed.
For example, if the 95~GHz emission is linked with the disk corona, it could manifest itself in a different X-ray/radio connection than that of RL AGN, where a relativistic jet plays a dominant role in the radio emission.

\begin{figure*}
\centerline{
\hskip -1cm
\includegraphics[scale=0.4,angle=90]{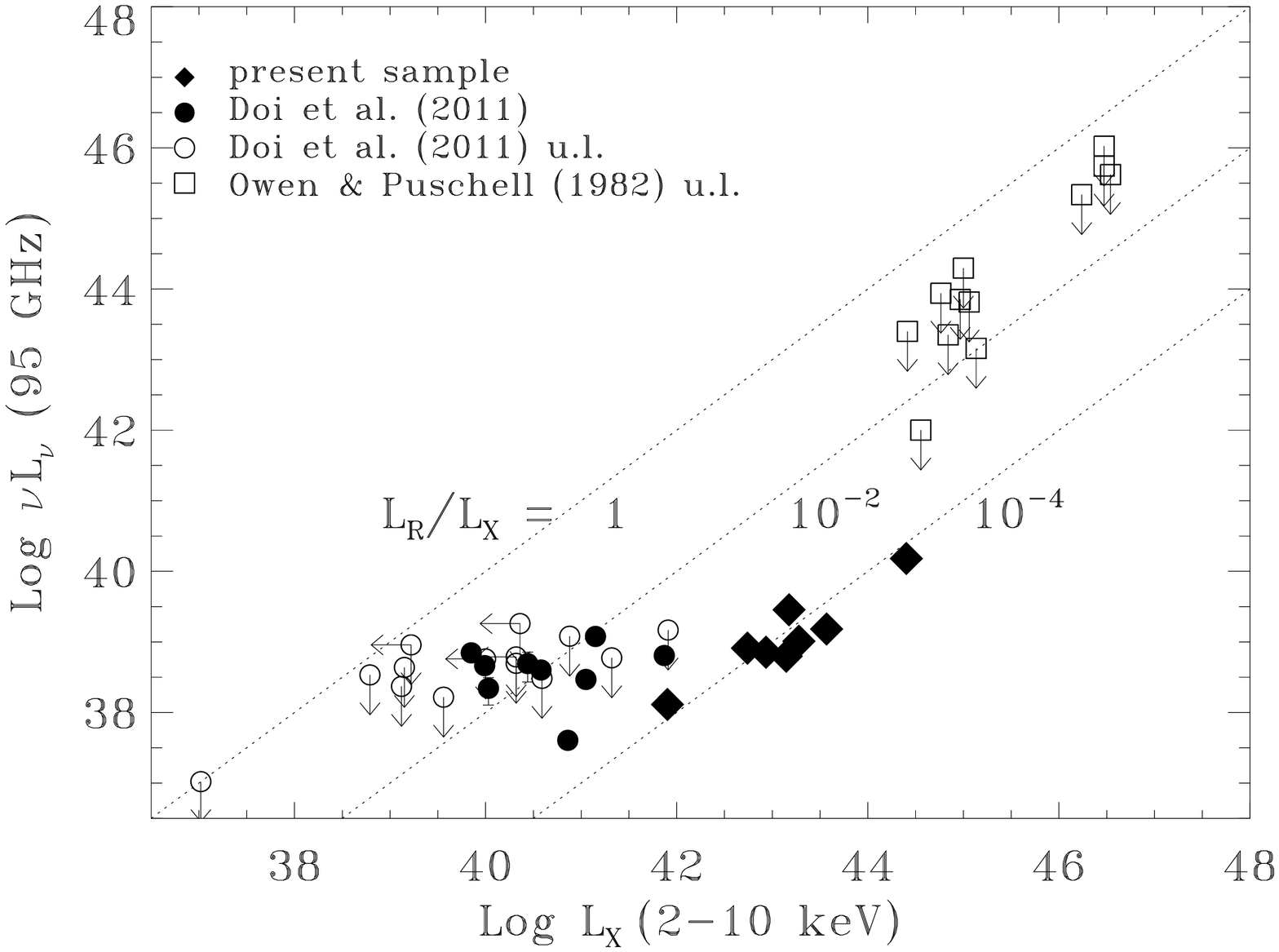}
\hskip -1cm
\includegraphics[scale=0.4,angle=90]{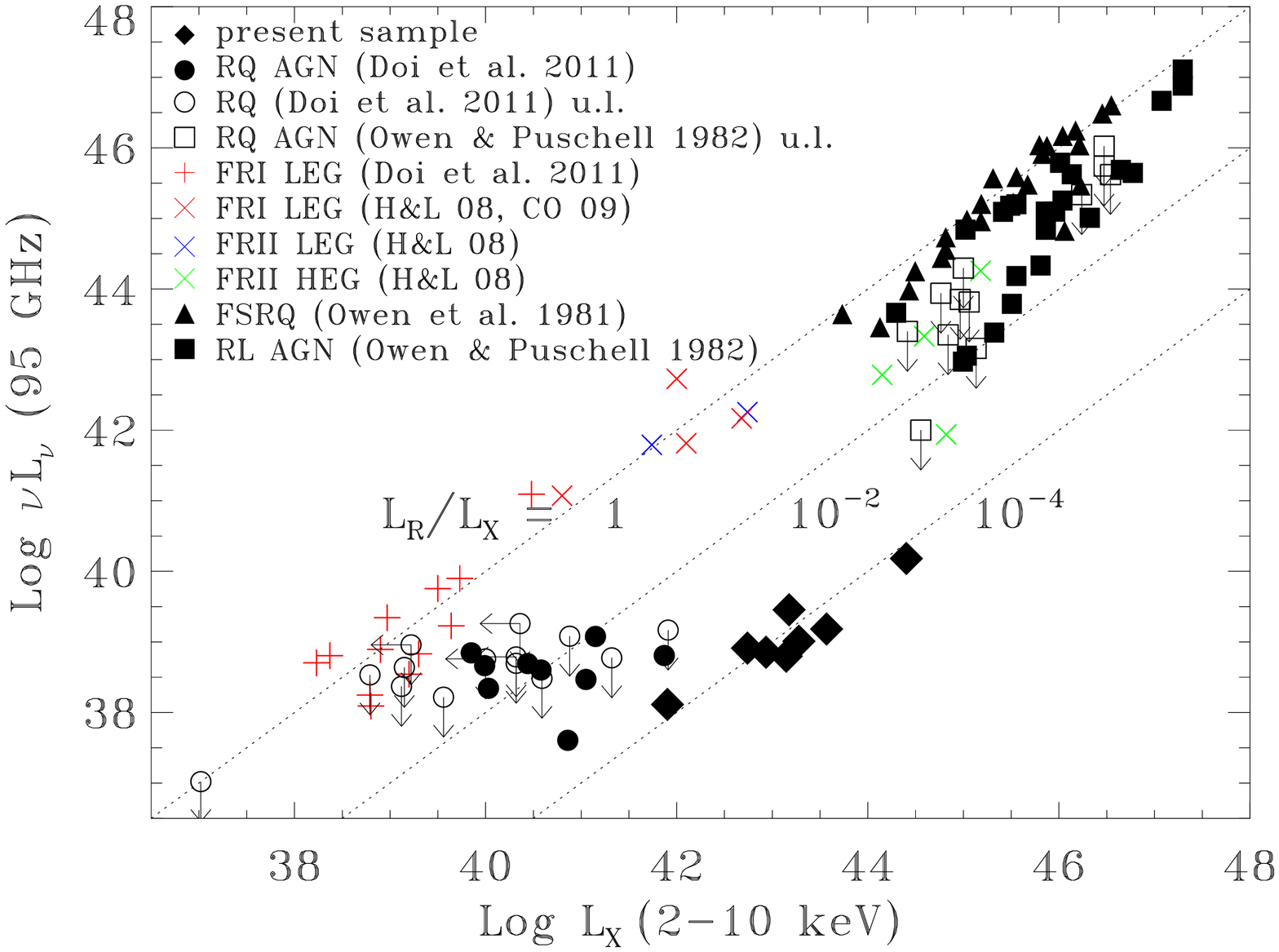}}
\caption{$L_R (\equiv \nu L_\nu)$ at 95~GHz vs. $L_X$ (2--10 keV) in erg s$^{-1}$ comparing the present sample (diamonds) with published observations of RQ ({\it left} and {\it right}) and RL ({\it right}) AGN cores. 
For RL sources, only detections are plotted.
The present sample is more X-ray luminous than that of \citet{doi11} and tightly follows the $L_R = 10^{-4} L_X$ correlation, while the RL LEG sources are spread around $L_R = L_X$. 
}
\label{LxL95}
\end{figure*}

Starting with RQ AGN, Figure~\ref{LxL95} (left panel) presents the comparison between the 95-GHz luminosity and the 2--10 keV X-ray luminosity
for the present sample, the RQ sources of \citet{doi11}, and the upper limits from \citet{owen82}, which includes much brighter sources.
We determined the RQ sources from \citet{doi11} based on their core 5 GHz luminosity and the associated $\log L_R /L_X <  -2.8$  criterion \citep{panessa07}\footnote{According to this criterion, the RQ AGN are NGC~404,
NGC~2768, NGC~2974, NGC~3065, NGC~3156, NGC~3610, NGC~4233, NGC~4494,
NGC~4742, NGC~5173, NGC~5283, NGC~5812, NGC~266, NGC~3147, NGC~3169,
NGC~3226, NGC~3718, NGC~4203, NGC~4143, NGC~4258, NGC~4565, NGC~4579,
and NGC~4772. The rest of the \citet{doi11} sample are RL AGN.}.
In the figure, we plot only 95~GHz flux densities, and average over multi-epoch data when present. 
More specifically, for NGC\,4143 the plotted mean flux density is 7.6 mJy, and for NGC\,4258 it is 8.6 mJy.
Many of the RQ targets in \citet{doi11} and all of those in  \citet{owen82} are upper limits. Detections are plotted as full symbols.

It can be seen that the detected sources in \citet{doi11} are concentrated below  $\log L_R \le 39$ ($L_R = \nu L_\nu$ in erg\,s$^{-1}$ at $\nu = 95$~GHz).
This is similar or only slightly less luminous than the present sources, except for MR\,2251-178 which is more luminous ($\log L_R = 40.18$).
Given that $\log L_R \le 39$  is about the sensitivity limit of small millimeter arrays observing nearby galaxies, and that many of the sources in \citet{doi11} have only upper limits, it is unclear how significant the difference at 95 GHz between the samples actually is.
The survey of \citet{owen82} is much less sensitive, and its RQ sources all have only upper limits at $\log L_R < 42$, 
so also consistent with all RQ AGN lying below $\log L_R \le 40$.
On the other hand, the sources of the present sample in X-rays lie between $42 < \log L_X < 44$, 
while those of \citet{doi11} occupy the $\log L_X < 42$ region of the scatter plot in Fig.~\ref{LxL95}, 
and those of \citet{owen82} are all X-ray bright with $\log L_X > 44$.

The eight present sources lie within a factor of 2 from the $L_R = 10^{-4}L_X$ relation.
This is reminiscent of the $L_R \sim 10^{-5}L_X$ relationship at 5 GHz \citep{laor08} that is also found in stars \citep{guedel93}.
Bear in mind that the X-ray luminosity used by \citet{laor08} was integrated over the 0.2--20 keV band, 
while in Fig.~\ref{LxL95} we take the 2--10 keV luminosities, as these are available in the literature.
Assuming a mean X-ray energy spectral slope of --1,
the present $L_X$ values would increase by a factor of $\sim$2.9 for the entire 0.2 -- 20~keV band. 
The frequency used here is 95~GHz, 
but for a radio spectral index of $\alpha = 1$, $L_R = \nu L_\nu$ is independent of frequency.
The relation adjusted for the extended X-ray band would  therefore yield on average $L_R \sim 3\times 10^{-5}L_X$.
Furthermore, X-ray absorbed quasars were excluded from the sample of \citet{laor08}, while most of the present AGNs are absorbed.
Accounting for X-ray absorption would bring the present $L_R - L_X$ relation even closer to the canonical $L_R \sim 10^{-5}L_X$ value,
which holds at 5~GHz.


The sources from the other samples in the left hand side of Fig.~\ref{LxL95} do not appear to lie along the $L_R = 10^{-4}L_X$ relation,
although many are upper limits.
The eight sources in the present sample were chosen based on their X-ray brightness and variability.
Hence, it is dangerous to reach a conclusion based only on their $L_R / L_X$ behavior and how it is different from other samples with other selection criteria.
Although their position on the $L_R / L_X$ diagram (Fig.~\ref{LxL95}) appears unique ($L_R / L_X \approx 10^{-4}$), this could easily be the result of the selection bias.
Indeed, sources from the sample of \citet{doi11} have lower X-ray luminosities, but radio luminosities comparable to the present sample.
On the other hand, the many upper limits impede a conclusive argument about the general $L_R / L_X$ behaviour. 

\subsubsection{Radio Loud Samples at 95~GHz}
Expanding the picture to include RL AGN, we plot in the right hand panel of Figure~\ref{LxL95}, 95 GHz and X-ray measurements of  FRI and FR~II low and high excitation galaxies (LEG and HEG) from \citet{doi11}, \citet[][CO in the figure]{cotton09}, and \citet[][H\&L in the figure]{hardcastle08},
who observed a small sample of powerful radio cores at 90 GHz. 
We also include the bright sample ($> 1$Jy) of RL quasars with flat or inverted spectra at millimeter wavelengths from \citet{owen81} and the mixed sample of RQ and RL quasars observed at 90 GHz by \citet{owen82}.
As anticipated, the RL sources occupy the upper envelope of the plotted distribution, 
which corresponds to higher radio luminosities at given $L_X$ values.

Among the RL AGN, a tight linear correlation is found for the LEG galaxies, both FR\,I and two FR\,II radio sources.
Considering the censored data, the relation is consistent with $L_R/L_X = 1.0 \pm 0.1$, with a very low probability of $< 0.0001$ that this is fortuitous. 
This relation complements similar relations found for LEG sources between the radio core power
(i.e., at 5~GHz) and multi-band nuclear luminosity, such as UV,
optical, infrared, and X-rays \citep[e.g.,][]{chiaberge99, chiaberge02, hardcastle00, balmaverde06, balmaverde06a, baldi10a}. 
These relations are interpreted as evidence for synchrotron emission from the basis of the jets at different frequencies. 
The $L_R/L_X = 10^{-4}$ values of the present sample are different by 4 orders of magnitude suggesting its underlying physical mechanism is completely different.
Having said that, the basis if the jet and an accretion disk corona are not necessarily mutually exclusive.

On the other hand, the RL HEG galaxies are brighter in the X-rays than the LEG by two orders of magnitude or more (Fig.~\ref{LxL95}). 
This X-ray emission is interpreted as coming from the accretion disk, in excess of the beamed component \citep[e.g.,][]{hardcastle00, dicken09, baldi10b},
and is the dominant X-ray source in RQ AGN. 
At yet higher 95~GHz luminosities, 
Fig.~\ref{LxL95} includes also the flat-spectrum radio quasars of \citet{owen81}, 
which are even slightly more powerful in the radio than those of \citet{owen82}. 

Although RL AGN are generally brighter at 95~GHz than the RQ AGN, the right hand panel of Fig.~\ref{LxL95}
lacks a clear bi-modality between RQ ($L_R/L_X \ll 1$) and RL ($L_R/L_X \approx 1$) AGN. 
Especially at low luminosities the difference between RL and RQ decreases, and the two samples appear to approach each other on the lower $L_R$ and $ L_X$ corner of Fig.~\ref{LxL95}.
This could be a result of the mixed and incomplete sample presented in the figure. 
On the other hand, if the black-hole mass is a crucial property for radio loudness, and high black-hole mass of $>$10$^{8-9}$~M$_{\odot}$ 
\citep{laor00, baldi10a, chiaberge11} is required to launch relativistic jets, then the continuous BH mass values in AGN could explain the continuous range of 95 GHz luminosities in the figure, where the RQ AGN, which generally do not exceed 10$^{40}$ erg\,s$^{-1}$ represent a 95~GHz disk contribution with no jets.
Large scale (kpc) radio jet emission tends to produce steep spectra, contributing most significantly at low frequencies.
The difference therefore between RL and RQ is more prominent at low frequencies than at 95~GHz. 

\section{Conclusions}
\label{sec:conclusions}

We observed eight X-ray bright and variable RQ AGN at 95~GHz with CARMA and ATCA.  All of the sources were detected.  
Millimeter-waves are the sole remaining unexplored band of AGN physics, and the present observations help to fill this band gap.
The emerging SED, based on archival data shows significant mm-wave emission from the inner core of the AGN, which dominates the spectrum at high radio frequencies.

The main features of the 95 GHz emission discovered in the present work can be summarized as follows:

\begin{itemize}
\item The observed sources show 95~GHz excess of up to $\times 7$ with respect to their extrapolated steep spectra at lower frequencies.
\item The deduced size $R_{\rm pc}$ for an optically thick 95~GHz radio-sphere is $10^{-4} - 10^{-3}$ pc, 
and in terms of gravitational radii ranges roughly from 10 -- 1000 $r_g$, similar in size to that of the X-ray source.
\item All eight sources lie tightly along an $L_R = 10^{-4}L_X$ correlation.
\end{itemize}

We interpret these findings as accumulating evidence for the 95~GHz emission being associated with the accretion disk, more specifically we suspect it originates near the X-ray corona.
The excess emission implies a compact, optically thick core.
The $L_R / L_X$ correlation is analogous to that found for RQ AGN and for stellar coronae at 5~GHz.
These are two separate lines of (circumstantial) evidence for an accretion disk/coronal origin 
of the compact 95~GHz source that is physically different from the majority of radio emission in RL AGN.

We suggest following up the present exploratory campaign of an X-ray selected sample with a systematic study of the SEDs of RQ AGN between 1 -- 100~GHz for a complete sample, selected independent of radio and X-ray properties.
For example, in the PG quasar sample this will allow to compare with the vast data available, 
and to explore possible dependence of the mm flux on the continuum and line emission properties in other bands.

In order to further investigate the conjecture of radio emission from
the accretion disk corona, where the X-rays originate,
simultaneous millimeter and X-ray observations are needed.
Temporal correlation between the X-ray and radio/millimeter light
curves from RQ AGN would be the smoking gun of coronal radio emission.
For example, a Neupert effect \citep{neupert68}, as first observed on the
Sun, and later in stellar coronae \citep[e.g.,][]{guedel02} is regarded as
direct evidence for chromospheric evaporation that produces X-ray
flares, as a result of coronal heating by non-thermal
electrons that produce the radio emission. 
This effect, or similar,
in AGN would constitute more evidence that the radio emission in RQ
AGN is of coronal origin, and linked to the X-ray emission from the
immediate vicinity of the accretion disk.

\section*{Acknowledgments}

We thank the referee for many helpful comments that led to improvements in the paper. 
We thank Robert (Ski) Antonucci for insightful comments.
This research is supported by the I-CORE program of the Planning and Budgeting Committee and the Israel Science Foundation (grant numbers 1937/12 and 1163/10), and by a grant from Israel's Ministry of Science and Technology. RDB was supported at the Technion by a fellowship from the Lady Davis Foundation.
This work was supported in part by the National Science Foundation under Grant No. PHYS-1066293 and the hospitality of the Aspen Center for Physics. 

The Australia Telescope is funded by the Commonwealth of Australia for operation as a National Facility managed by CSIRO.
Support for CARMA construction was derived from the states of California, Illinois, and Maryland, the James S. McDonnell Foundation, the Gordon and Betty Moore Foundation, the Kenneth T. and Eileen L. Norris Foundation, the University of Chicago, the Associates of the California Institute of Technology, and the National Science Foundation. 

\label{lastpage}

\bibliography{mn-jour,my}

\end{document}